\documentclass[aps,nofootinbib,preprint]{revtex4}
\pdfoutput=1
\usepackage{graphicx,subfigure}
\usepackage[bookmarks=true]{hyperref}
\usepackage{amsmath}
\usepackage{amsfonts}
\usepackage{amssymb}
\usepackage{slashed}

\usepackage{cprotect}

\usepackage{amsmath}
\usepackage{amsfonts}
\usepackage{amssymb}
\usepackage{color}

\def\GeV{\rm{GeV}}
\def\TeV{\rm{TeV}}

\def\be{\begin{equation}}
\def\ee{\end{equation}}
\def\bea{\begin{eqnarray}}
\def\eea{\end{eqnarray}}

\def\ss2l{SS2$\ell$}
\def\3l{3$\ell$}

\def\slashchar#1{\setbox0=\hbox{$#1$}              \dimen0=\wd0                                    \setbox1=\hbox{/} \dimen1=\wd1                  \ifdim\dimen0>\dimen1                              \rlap{\hbox to \dimen0{\hfil/\hfil}}            #1                                           \else                                              \rlap{\hbox to \dimen1{\hfil$#1$\hfil}}         /                                            \fi}

\begin{document}
\title{Usefulness of effective field theory for boosted Higgs production}
\vspace*{1cm}

\author{S.~Dawson}
\affiliation{Department of Physics, Brookhaven National Laboratory, Upton, N.Y., 11973,  U.S.A.}
\author{I.~M.~Lewis}
\affiliation{SLAC National Accelerator Laboratory, Menlo Park, CA, 94025, U.S.A.}
\author{Mao Zeng}
\affiliation{C.N. Yang Institute for Theoretical Physics, Stony Brook University, Stony Brook, N.Y., 11794, U.S.A.}

\begin{abstract}
\vspace*{0.5cm}
The Higgs + jet channel at the LHC is sensitive to the effects of new physics 
both in the total rate and in the transverse momentum distribution 
at high $p_T$.  We examine the production process
using an effective field theory (EFT) language and discuss the possibility of determining the nature of the 
underlying high
scale physics from boosted Higgs production.  The effects of heavy color
triplet  scalars and top partner fermions
with  TeV scale masses are considered as examples and Higgs-gluon couplings of dimension-5 and 
dimension-7 are included in the EFT. As a by-product of our study, we examine the region of validity of the EFT.
Dimension-7 contributions in realistic new physics models give effects in the high $p_T$ tail of the Higgs signal which are so tiny that they are likely to 
be unobservable.

\end{abstract}

\maketitle

\section{Introduction}

\makeatletter{}The recently discovered Higgs boson has all the generic characteristics of a  Standard Model (SM)
 Higgs boson and
measurements of the production and decay rates
agree to the $\sim 20\%$ level with Standard Model
 predictions \cite{ATLAS-CONF-2014-009,CMS-PAS-HIG-14-009,Dittmaier:2011ti,Dittmaier:2012vm}.  
 Precision measurements of Higgs couplings are essential for understanding whether
 there exist small
  deviations
 from the Standard Model predictions which  could be  indications of undiscovered high scale physics.
 If there are no  weak scale particles beyond those of the SM, then effective field theory (EFT)  techniques
 can be used to probe the Beyond the Standard Model  (BSM)
  physics \cite{Hagiwara:1993ck,Alam:1997nk,Henning:2014wua}.  
 The EFT is the most general description of low energy processes 
 and new physics manifests itself  as
 small deviations from the SM predictions. In the electroweak sector, this approach has been 
 extensively studied \cite{Giudice:2007fh,Corbett:2012dm,Falkowski:2014tna,Ellis:2014dva,Chen:2013kfa}.
 The  effects of BSM operators affecting Higgs production in the strong sector have been less studied \cite{Dawson:2014ora,Manohar:2006gz,Neumann:2014nha}.

  The largest contribution to Standard Model
Higgs boson production at the LHC comes from gluon fusion through a top quark loop and we examine new physics effects in this channel, along with the related Higgs + jet channel. 
 We consider
an effective Lagrangian containing the SM fermions and gauge bosons, along with a single
Higgs boson, $h$.
At dimension-4,
the fermion- Higgs couplings can be altered from the SM couplings by a simple rescaling,
\begin{equation}
-\mathcal L_f={\kappa_f}\biggl({m_f\over v}\biggr) {\overline f} f h +H.c.\, ,
\label{eq:kappaF}
\end{equation}
where $\kappa_f=1$ in the SM.
In models with new physics, 
the gluon fusion rate can also  be altered by new heavy particles interacting
with the Higgs boson at one-loop,  which contribute to an effective dimension-5 
operator \cite{Kniehl:1995tn,Spira:1995rr,Dawson:1990zj}
\begin{equation}
\mathcal L_5=C_1 G^{A,\mu\nu}G_{\mu\nu}^A h\, ,
\label{l5def}
\end{equation}
where $C_1=\alpha_s / (12 \pi v) $ 
for an infinitely heavy fermion with $\kappa_f=1$.
For convenience, we define $\kappa_g$ to be the ratio of $C_1$ to this reference value,
\begin{equation}
\kappa_g \equiv C_1 / \left( \frac{\alpha_s}{12\pi v} \right).
\label{eq:defKg}
\end{equation}

\begin{figure}[t] 
\begin{centering}
\includegraphics[scale=0.5]{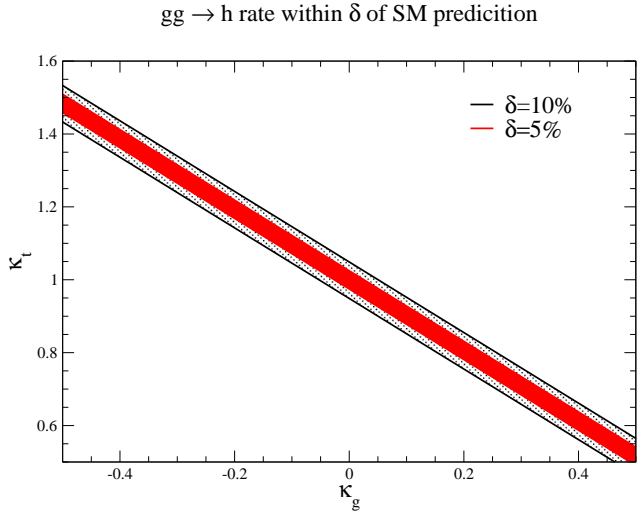}
\par\end{centering}
\caption{Allowed values of the EFT coefficients when the total gluon fusion rate, $gg\rightarrow h$, 
is within $\pm 10\% (\pm 5\%)$ of the SM prediction, ($\kappa_g\equiv 12\pi v C_1 / \alpha_s$) . }
\label{fig:total}
\end{figure}

We compute the top quark contribution to scattering processes exactly using Eq. \ref{eq:kappaF}, (${\it {i.e.}}$, not
in the infinite top quark mass limit),
and consider $C_1$ to be only the contribution from new physics. 
The measurement of gluon fusion by itself can determine a combination 
of $\kappa_g$ and the top quark Yukawa coupling, $\kappa_t$, 
but cannot distinguish between the two for $m_t \gg m_h$
 \cite{Low:2009di,Grojean:2013nya,Azatov:2013xha,Banfi:2013yoa}. 
 Including  the dimension-5 operator of Eq. \eqref{l5def},
  the cross section
 is generically,
 \begin{equation}
 \mu_{ggh}\equiv {\sigma(gg\rightarrow h)\over \sigma(gg\rightarrow h)_{SM}}\sim
 \mid \kappa_t + \kappa_g \mid^2+{\cal O}\biggl({m_h^2\over m_t^2}\biggr)\, .
 \label{eq:SMrel}
 \end{equation}
 The requirement that $\mid \mu_{ggh}-1\mid < 10\%$ (or $5\%$) 
 is shown in Fig. \ref{fig:total},
 where top quark mass effects are included exactly.
    The SM corresponds to the point $\kappa_g=0, \kappa_t=1$.  The contribution from $b-$ quarks is small and has been neglected. 

The boosted production of the Higgs boson through the process 
$pp\rightarrow h$+jet is sensitive  to the 
Higgs-  
gluon effective coupling \cite{Banfi:2013yoa,Grojean:2013nya,Buschmann:2014twa,Azatov:2013xha,Schlaffer:2014osa,Buschmann:2014sia} and offers the possibility  of disentangling new physics effects  and hence breaking the degeneracy between $\kappa_t$ and
$\kappa_g$.   An effective Lagrangian approach is useful for studying this high $p_T$ BSM physics   and 
the  Higgs-parton interactions
can be
described as a sum of higher dimension operators,
\begin{equation}
\mathcal L_{EFT}\sim \mathcal L_4+ \mathcal L_5+\mathcal L_6+ \mathcal L_7+ \dots\, ,
\end{equation}
where $\mathcal L_n$ includes all dimension-$n$
operators.  At dimension-5 and assuming CP conservation, there is only the single operator
of Eq. \eqref{l5def} modifying the Higgs-gluon interactions.  The dimension-5 operator has been broadly
used to obtain higher order QCD corrections to Higgs rates \cite{Dawson:1990zj,Spira:1995rr,Harlander:2002wh,Anastasiou:2002yz,Glosser:2002gm,deFlorian:1999zd,Ravindran:2003um,Ravindran:2002dc,Boughezal:2013uia}.

Dimension-7 operators affecting Higgs- gluon interactions from QCD interactions have received less
 attention \cite{Neill:2009tn,Harlander:2013oja,Manohar:2006gz}. Because their
 contributions are proportional
to the strong coupling, $g_s$, these operators can have numerically significant effects.
In a previous work \cite{Dawson:2014ora},
 we considered the effects of dimension-7 operators
 affecting Higgs- gluon interactions
  and 
 demonstrated the importance of including these operators along with the
  NLO QCD corrections in order to obtain realistic predictions
 of boosted Higgs spectra.   
 The largest contribution to Higgs + jet production is from the $O_1$ operators in the $gg$ initial channel.
 The NLO QCD corrections to this channel are relatively flat in $p_T$ and lead to an enhancement of roughly
 a factor of $2$ in the rate at the $14$ TeV LHC.
 The contributions from $O_3$ to Higgs + jet production are suppressed
 at lowest order QCD (LO) for large $p_T$, since they vanish in the soft Higgs limit.  These contributions receive
 large NLO corrections, but remain numerically small and are never important.  The contributions
  from the interference of the $O_1$ and $O_5$ operators can be important for large $p_T\sim 300~GeV$
and receive NLO QCD corrections which are again fairly $p_T$ independent and increase the rate by
a factor of $\sim 1.2$.

  In this paper, we
examine the expected size of the coefficients of  the Higgs-gluon EFT dimension-5
and dimension-7 operators in several representative UV  models with heavy colored
scalars and fermions.
We are particularly interested in the question of whether the measurement  of the
boosted Higgs $p_T$ distribution
can distinguish the nature of the underlying UV physics, should there be any deviation from the SM.  We then demonstrate how the inclusion of the dimension-7
operators affects fit to EFT Higgs parameters from gluon fusion.  We work at LO QCD.

In Section \ref{efflag}, we review the EFT.
The heavy colored scalar and fermion models which we study are introduced in Section \ref{bsm} and the matching
coefficients of the EFT presented.  Phenomenological results at the LHC are given in Section IV  and some conclusions
about the usefulness of the EFT in this channel presented in Section \ref{conc}.

\section{Effective Lagrangian}
\label{efflag}
\makeatletter{}In this section, we review the effective Lagrangian relevant for Higgs + jet production containing
non-SM Higgs-gluon interactions.
We consider a CP conserving Lagrangian, with no new Higgs particles, 
\begin{equation}
\mathcal L = \mathcal L_{SM} + (\kappa_t - 1) (-1) \bar t t h + \mathcal L_5 + \mathcal L_7 + \dots,
\end{equation}
where
\begin{equation}
\mathcal L_5+ \mathcal L_7\equiv  \hat C_1 O_1+\Sigma_{i=2,3,4,5} \hat C_i O_i
\, ,\label{ldef}
\end{equation}
Note that there are no relevant dimension-$6$ operators of the type we are considering.

At dimension-$5$, the unique operator is
\begin{equation}
O_1=G_{\mu\nu}^A G^{\mu\nu,A}h\, ,
\end{equation}
where $G^A_{\mu\nu}$ is the gluon field strength tensor.  
The dimension-$7$ operators needed for the  gluon fusion production of Higgs are \cite{Neill:2009tn,Harlander:2013oja,Buchmuller:1985jz},
\begin{eqnarray}
O_2&=&  D_\sigma G^A_{\mu\nu}D^\sigma G^{A,\mu\nu}h \label{op2}
\\
O_3&=&f_{ABC}G_\nu^{A,\mu} G_\sigma^{B,\nu}G_\mu^{C,\sigma} h \label{op3}
\\
O_4&=&g_s^2h \Sigma_{i,j=1}^{n_{lf}} {\overline \psi}_i\gamma_\mu T^A \psi_i \,
 {\overline \psi}_j\gamma^\mu T^A \psi_j \label{op4}
\\
O_5&=& g_s h\Sigma_{i=1}^{n_{lf}} 
G_{\mu\nu}^A D^\mu\, {\overline \psi}_i\gamma^\nu T^A\psi_i\, ,
\label{op5}
\end{eqnarray}
where our convention for the covariant derivative is $D^\sigma=\partial^\sigma -ig_s T^A G^{A,\sigma}  $,
$Tr(T^AT^B)={1\over 2}\delta_{AB}$
and $n_{lf}=5$ is the number of light fermions.
 Including  light quarks,  $O_4$ and $O_5$ are needed, which are related by the equations of motion
 (eom)  to gluon-Higgs operators, 
 \begin{eqnarray}
 O_4\mid_{eom}&\rightarrow &  D^\sigma G^A_{\sigma\nu}D_\rho G^{A,\rho\nu} h \equiv O_4'
 \nonumber \label{op4prime} \\
 O_5\mid_{eom} &\rightarrow & G^A_{\sigma\nu}D^\nu D^\rho G_\rho^{A,\sigma} h \equiv O_5' \, .
\label{op5prime}
\end{eqnarray} 
Since $O_4$ involves 4 light fermions, the operator only contributes to Higgs + jet production starting at NLO.

A different dimension- $7$ operator is useful,
\begin{equation}
O_6 = -D^{\rho} D_{\rho} \left( G_{\mu\nu}^A G^{\mu\nu,A} \right) h = 
-\partial^{\rho} \partial _{\rho} \left( G_{\mu\nu}^A G^{\mu\nu,A} \right) h = 
m_h^2 O_1,
\label{opRelation}
\end{equation}
where the last equal sign is only valid for on-shell Higgs production.
  Using the Jacobi identities, 
\begin{equation}
O_6 = m_h^2 O_1 = -2 O_2 + 4 g_s O_3 + 4 O_5.
\label{oprel}
\end{equation}
Therefore, we can choose $O_6 = m_h^2 O_1$, $O_3$, $O_4$, and $O_5$ as a complete basis for 
the dimension- $7$ Higgs-gluon-light quark operators. We  rewrite Eq. \eqref{ldef} as
\begin{equation}
\mathcal L_{\rm eff} = C_1 O_1 +  \left( C_3 O_3 + C_4 O_4 + C_5 O_5 \right). \label{ldef1}
\end{equation}

The lowest order amplitudes for Higgs + jet production including all fermion mass dependence (bottom and top) are given in
Refs. \cite{Baur:1989cm,Ellis:1987xu}. 
A study of Higgs + jet production at LO QCD in the EFT approximation involves 
only $C_1,C_3$ and $C_5$ \cite{Dawson:2014ora,Neill:2009mz}.
At the lowest order in $\alpha_s$, $O_3$ is the only dimension-7 operator which contributes to the $gg \rightarrow gh$ channel, while $O_5$ is the only dimension-$7$ operator which contributes to channels with initial state quarks.
The lowest order amplitudes in the EFT for Higgs + jet production can be found in Ref. \cite{Dawson:2014ora}, 
along with the NLO
results including the effects of dimension-$7$ operators.
For Higgs + jet production at NLO in BSM models, the EFT description also needs to include the higher-dimensional $3-$gluon effective vertex generated at one-loop \cite{Dawson:2014ora,Ghosh:2014wxa}, which could affect dijet and top quark rates \cite{Cho:1994yu}.   
 
\section{UV Physics AND THE EFT}
\label{bsm}
\makeatletter{}In this section, we discuss several prototype BSM physics models which have heavy particles contributing to Higgs + jet production
and we compute the matching coefficients for the EFT in these models.  This will allow us to estimate the
size of BSM contributions.  
\subsection{Heavy Colored Scalars}
We consider the addition of either real or complex $SU(3)$ scalars, $\phi_i$ \cite{Bonciani:2007ex,Arnesen:2008fb,Kribs:2012kz,Boughezal:2010ry,Gori:2013mia}.   Our numerical results are all derived for a complex
scalar triplet.
The scalar portion of the Lagrangian 
involving a  new complex scalar, $\phi_i$,   and the SM-like Higgs doublet, $H$, is ,
\begin{eqnarray}
V_{\rm complex}&=&V_{SM}(H)+m_i^2\phi_i^\dagger \phi_i 
+{C_h\over v}\phi_i^\dagger \phi_i (H^\dagger H)  -\lambda_4(\phi_i^\dagger\phi_i)^2
\, ,
\end{eqnarray}
where $V_{SM}$ is the  SM Higgs potential.
For a real scalar, 
\begin{eqnarray}
V_{\rm real}&=& 
V_{SM}
+{m_i^2\over 2} (\phi_i)^2
+{C_h\over 2v}(\phi_i)^2 (H^\dagger H)  -\lambda_4(\phi_i)^4
\, .
\end{eqnarray}
In unitary gauge,   $H\rightarrow (0, (h+v)/\sqrt{2})$.
 
\subsection{Top Partner Model}

Many BSM
contain a charge - ${2\over 3} $ partner of the top quark.
 We consider a general case
with a vector-like $SU(2)_L$ singlet fermion which is allowed
to mix with the Standard Model like top 
quark \cite{ Lavoura:1992qd,AguilarSaavedra:2002kr,Aguilar-Saavedra:2013qpa,
Popovic:2000dx,Dawson:2012di}.
The mass eigenstates are defined to be $t$ and $T$ with masses $m_t$ and $M_T$ 
and  are derived from the gauge eigenstates using
bi-unitary transformations involving two mixing angles $\theta_L$ and $\theta_R$.
Without loss of generality, $\theta_R$ can be removed by a redefinition of the top partner
gauge eigenstate and the Higgs couplings are then modified from 
those of
the SM \cite{Dawson:2012mk}:
\begin{eqnarray}
L_h^{top~partner}&=&-\biggl\{ \cos^2\theta_L{m_t\over v} {\overline t}_L t_Rh
+\sin^2\theta_L{M_T\over v} {\overline T}_L T_R h\nonumber \\ &&
+{M_T\over 2v}\sin(2\theta_L) {\overline t}_LT_Rh+{m_t\over2 v} \sin(2\theta_L){\overline T}_L t_R h
+H.c.\biggr\}\, .\label{eq:TopPartner} \end{eqnarray}
Precision electroweak fits to the oblique parameters, as well as $M_W$,
 place stringent restrictions on the product $\sin^2\theta_L M_T^2$ and
for $M_T\sim 1~\TeV$, $\sin \theta_L < .17$ \cite{Dawson:2012di,Aguilar-Saavedra:2013qpa}.
Higgs production has been investigated at NNLO for top partner models in Ref. \cite{Dawson:2012di}
and the rate determined to be within a few $\%$ of the SM rate for allowed values of $\theta_L$. Large effects in this
channel require values of $\sin\theta_L$ that are excluded by precision measurements. 
ATLAS \cite{Aad:2014efa}  and CMS \cite{Chatrchyan:2013uxa} have searched for top singlet partners and excluded $M_T$ below $655$~\GeV
and $687$~\GeV, respectively.  Similar limits on top partner masses and mixing can be obtained for different
choices of top partner $SU(2)_L$ properties \cite{Aguilar-Saavedra:2013qpa}.

\subsection{Predictions for Coefficients}

The exact results for the contributions from  high scale fermion \cite{Baur:1989cm,Ellis:1987xu} and scalar 
loops \cite{Bonciani:2007ex,Arnesen:2008fb} to the rates for $q {\overline q}
\rightarrow g h$ and $gg\rightarrow
gh$ are well known.  Matching to the EFT expressions,
the coefficient functions can be extracted. 
The EFT amplitude for $q{\overline q} \rightarrow gh$ from virtual heavy particles with mass, $m$, is
\begin{eqnarray}
\mid A(q {\overline q}\rightarrow gh)\mid^2 &=& 
64 g_s^2 \biggl(
{{\hat t}^2+{\hat u}^2\over {\hat s}}
\biggr)
\biggl[ C_1^2
+{{\hat s} C_1 C_5\over 2}\biggr]\nonumber \\
&=& \lim_{ m\rightarrow\infty}
\biggl(
{4\alpha_s^3\over \pi}\biggr)
\biggl({{\hat u}^2+{\hat t} ^2\over {\hat s}v^2}\biggr)
\mid {\cal{A}}_5({\hat s},{\hat t},{\hat u},m^2)\mid^2\, ,
\label{match1}
\end{eqnarray}
while 
the EFT amplitude for $gg\rightarrow gh$ from virtual heavy particles with mass, $m$, is
\begin{eqnarray}
\mid A(gg\rightarrow gh)\mid^2 &=& 
g_s^2\biggl[384 C_1^2 \biggl[
{m_h^8+{\hat s}^4+{\hat t}^4+{\hat u}^4\over {\hat s}{\hat t}{\hat u}}\biggr]
+1152 C_1 C_3 m_h^4 \biggr]\nonumber \\
&=& \lim_{ m\rightarrow\infty}
\biggl({96\alpha_s^3\over \pi}{m_h^8\over {\hat s}{\hat t}{\hat u}v^2}\biggr) \biggl\{
\mid {\cal{A}}_2({\hat s},{\hat t},{\hat u},m^2)\mid^2+
\mid {\cal{A}}_2({\hat u},{\hat s},{\hat t},m^2)\mid^2
\nonumber \\
&&+\mid {\cal{A}}_2({\hat t},{\hat u},{\hat s},m^2)\mid^2+
\mid {\cal{A}}_4({\hat s},{\hat t},{\hat u},m^2)\mid^2\biggr\}\, ,
\label{match2}
\end{eqnarray}
where ${\hat s},{\hat t},$ and ${\hat u}$ are the usual Mandelstam variables. 
The coefficient functions ${\cal{A}}_2({\hat s},{\hat t},{\hat u},m^2)$, 
${\cal{A}}_4({\hat s},{\hat t},{\hat u},m^2)$
and ${\cal {A}}_5({\hat s},{\hat t},{\hat u},m^2)$  are given in  Ref. \cite{Ellis:1987xu} for fermion loops and in 
Ref. \cite{Bonciani:2007ex} for scalar loops.  
The $C_1, C_3$ and $C_5$ coefficients of Eqs. \ref{match1} and \ref{match2} depend in general on the parameters of the underlying
UV completion of the model. 
By matching
the EFT predictions with the heavy fermion expansions, we obtain the EFT coefficients given
in Table \ref{tabcof}.  At LO, the dimension -$7$ term contributing to  the $gg\rightarrow gh$ amplitude does not contain any dependence
on the kinematic variables.  For $\TeV$ scale masses, it is clear that the coefficients are quite small.
For the top partner model, the coefficient functions for the heavier Dirac fermion contributions need to be multiplied by the factor $\sin^2 \theta_L$ appearing in Eq. \eqref{eq:TopPartner}, while the SM top quark contribution is included exactly without using the EFT.

\begin{table}[tp]
\centering
\begin{tabular}{|c|c|c|c|}
\hline
  & Dirac Fermion & $SU(3)$ Triplet Scalar & $SU(3)$ Octet Scalar \\
  \hline
  $C_1(\Lambda)$ & ${\alpha_s \kappa_F\over 12\pi v}\biggl[1+{7m_h^2\over 120 m_F^2}\biggr]$
  &$-{\alpha_s\over 96 \pi M_S^2 } C_h \biggl[1+{2m_h^2\over 15 M_S^2}\biggr]$ & 
  $-{\alpha_s\over 16 \pi M_S^2 } C_h \biggl[1+{2m_h^2\over 15 M_S^2}\biggr]$ \\
  $C_3(\Lambda)$ & $-{g_s \alpha_s \kappa_F\over 360 \pi v m_F^2}$ &
  $-{g_s\alpha_s\over 1440M_S^4}C_h$
  &$-{g_s\alpha_s\over 240M_S^4}C_h$
 \\
  $C_5(\Lambda)$ & ${11 \kappa_F\alpha_s \over 360 \pi v m_F^2}$ &
  $-{\alpha_s\over 360 \pi M_S^4}C_h$
  &$-{\alpha_s\over 60 \pi M_S^4}C_h$
  \\ 
\hline
\end{tabular}
\vspace{-1ex}
\caption{The effective Lagrangian coefficient functions for heavy Dirac fermions and heavy scalars
with mass, $m_F$ and $M_S$, respectively. 
The coefficient functions, along with $g_s$ and $\alpha_s$, are evaluated at the scale $\Lambda=m_{F}, M_S$.}
\label{tabcof}
\end{table}

\begin{figure}[t]
      \includegraphics[width=0.6\textwidth,angle=0,clip]{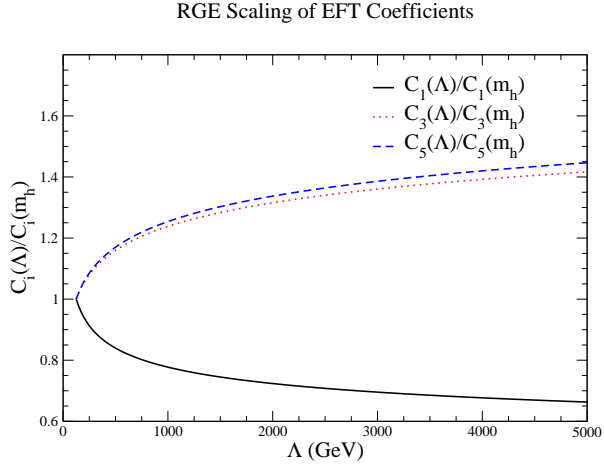}
\caption{The evolution of the dimension- $5$ and dimension-$7$ EFT coefficients from the scale of
new physics, $\sim \Lambda$, to the electroweak scale. }
\label{hgg_fig}
\end{figure}

The matching of the EFT and the underlying UV theory are done
at the high scale  $\Lambda$.  Using the anomalous dimensions found in Ref. \cite{Dawson:2014ora, Gracey:2002he},
the coefficients can be evolved to a low scale, $\mu_R\sim m_h$,
\begin{align}
\frac {d} {d \ln \mu_R} \ln \left( \frac {C_1(\mu_R)}{g_s^2(\mu_R)}
 \right) &= \mathcal O(\alpha_s^2(\mu_R)), \label{c1run} \\
\frac {d} {d \ln \mu_R} \ln \left( \frac {C_3(\mu_R)}{g_s^3(\mu_R)}
 \right) &= \frac{\alpha_s(\mu_R)}{\pi}\, 3 C_A, \label{c3run} \\
\frac {d} {d \ln \mu_R} \ln \left( \frac {C_5(\mu_R)}{g_s^2(\mu_R)} \right) &= \frac{\alpha_s(\mu_R)}{\pi} \left( \frac {11}{6}C_A + \frac{4}{3} C_F \right)\, , \label{c5run}
\end{align}
where $C_A=3$ and $C_F={4\over 3}$. The one-loop electroweak RG running of $C_1 / g_s^2$ \cite{Grojean:2013kd} is non-zero, and its effect on the Higgs $p_T$ distribution in the TeV range is found to be at the percent level \cite{Englert:2014cva}.

The leading-logarithmic solutions to the renormalization group running equations Eq. \eqref{c1run}-\eqref{c5run} are
\begin{align}
C_1(\mu_R) / g_s^2 (\mu_R) &= C_1(\mu_0) / g_s^2 (\mu_0), \\
C_3(\mu_R) / g_s^3 (\mu_R) &= \left( \frac{\alpha_s(\mu_R)}{\alpha_s(\mu_0)} \right)^{-\frac {3 C_A}{2b_0}} \cdot C_3(\mu_0) / g_s^3 (\mu_0), \\
C_5(\mu_R) / g_s^2 (\mu_R) &= \left( \frac{\alpha_s(\mu_R)}{\alpha_s(\mu_0)} \right)^{-\frac {1}{2b_0} \left( \frac{11}{6} C_A + \frac 4 3 C_F \right)} \cdot C_5(\mu_0) / g_s^2 (\mu_0)\, ,
\end{align}
where $b_0={1\over 12}(11C_A-2n_{lf})$
and $\mu_0\sim\Lambda$.  The evolution of the coefficient functions is shown in Fig. \ref{hgg_fig}.  $C_1$ is increased by $\sim$ a 
factor of $2$ when evolving from $\Lambda\sim 5~\TeV$ to the weak scale, while $C_3$ and $C_5$ are reduced by a similar factor.

\section{Phenomenology}
\label{phenosec}
\makeatletter{}
We will eventually be interested
in whether measurements of the $p_T$ spectrum can distinguish between the 
effects of the 
dimension-5 and dimension-7 operators resulting from scalars and from fermions; that is,  
\emph{``Is the EFT a
useful tool for disentangling the source of high scale physics?"}

Throughout this paper, diagrams involving the SM top quark are evaluated with exact $m_t$ dependence without using the Higgs-gluon EFT, while the contributions from heavy BSM particles, such as a color triplet scalar or a fermionic top partner, are considered both exactly and in the EFT approximation.
\subsection{Heavy Colored Scalars}
\label{subsec:scalar}
We begin by considering  the effect of heavy color triplet scalars on Higgs + jet production. (The case of a light colored scalar has been considered in \cite{Arnesen:2008fb}.)
We use CJ12 NLO PDFs \cite{Owens:2012bv} and $\mu_R=\mu_F=\sqrt{m_h^2+p_T^2}$
for all curves, with $m_h=125~\GeV$, $m_t=173~\GeV$, and $m_b=4.5~\GeV$.   All plots refer
to Higgs + jet production at lowest order and with $\sqrt{s}=14~$\TeV.
When using the EFT, the effects of heavy scalars are included using  the coefficients of Table I. Since
the effects are suppressed by $1/M_S^2$ in $C_1$ and
$1/M_S^4$ in the other $C_i$, we expect relatively small effects unless the coefficient
function $C_h$ is large.  We expect $C_h$ to be of order the electroweak scale in a realistic
model and  in our plots, we take $C_h=3M_Z$.\footnote{If $\phi_i$ corresponds to the left-handed top squark of the MSSM,
then in the alignment limit ($\sin\beta=\cos\alpha$), $C_h\sim 3 M_Z$, which motivates our choice. This numerical value is
not important for our conclusions, as long as $C_h/M_Z$ is not a large number.} Numerically, the effects are linear in $C_h$ for
modest values of $C_h / M_Z$ and our results 
can be trivially rescaled.   

\begin{figure}[t]
\begin{centering}
\includegraphics[scale=0.38]{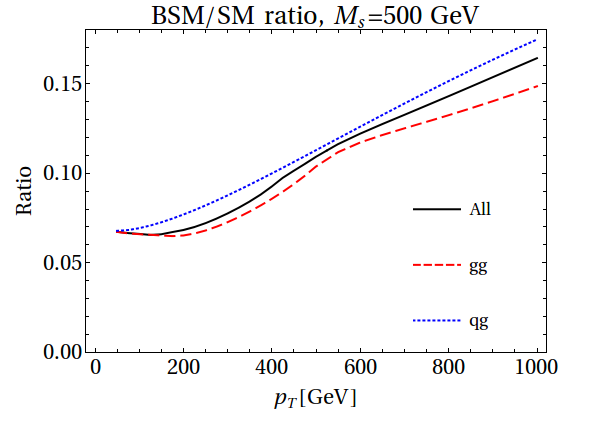}
\includegraphics[scale=.38]{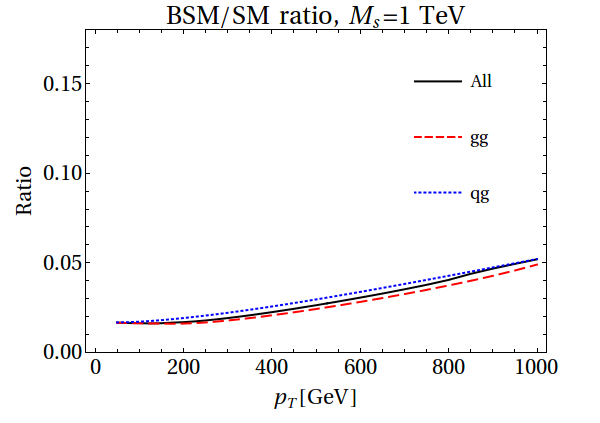}
\par\end{centering}
\caption{Contribution of a $500~\GeV$ color triplet scalar (LHS) and a $1~\TeV$ scalar (RHS),
 relative to the SM Higgs $p_T$ distribution. The $gg$ and $qg$ partonic channels, and the sum of all partonic channels (which also includes $q\bar q$), are shown separately. Both the SM top and the scalar contributions
 are included exactly at LO.} 
  \label{fig:scal_tot}
\end{figure}

\begin{figure}[t]
\begin{centering}
\includegraphics[scale=0.38]{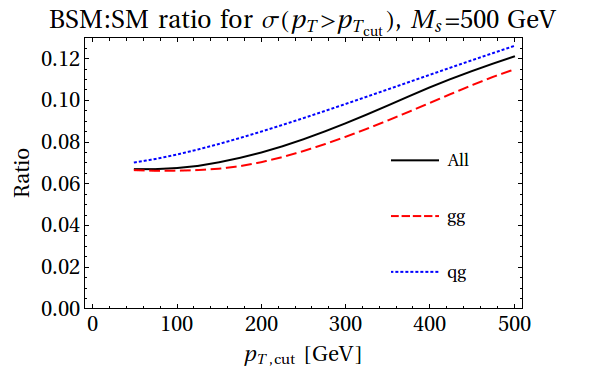}
\includegraphics[scale=.38]{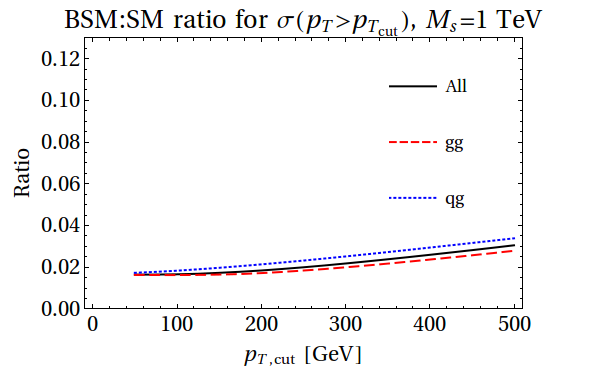}
\par\end{centering}
\caption{Contribution of a $500~\GeV$ color triplet scalar and a $1~\TeV$ scalar,
 relative to the SM cross section, with
 a cut $p_{T_{cut}}$. The $gg$ and $qg$ partonic channels, and the sum of all partonic channels (which also include $q\bar q$), are shown separately.Both the SM top and the scalar contributions
 are included exactly at LO.} 
\label{fig:scal_cut}
\end{figure}

The exact one-loop contribution of the heavy scalars  relative to the SM rate are shown
in Fig. \ref{fig:scal_tot}  and as expected, they cause 
only a few percent deviation from the SM rate at low $p_T$.  
We define the ratio, ``BSM/SM'' to be the differential (or integrated) rate in the theory with
the SM top quark and scalar included exactly normalized to the SM rate minus $1$, i.e. it is
the incremental contribution from the addition of a scalar. 
At large $p_T$, the deviation becomes significant,
approaching $\sim 15\%$ for $p_T\sim 1~\TeV$ for a $500~\GeV$ scalar and $\sim 5\%$ for a $1~\TeV$
scalar.  
We note that the effects of a color octet scalar are a factor of $C_A / T_F = 6$ larger than those of a color triplet scalar. 
The integrated cross sections with a $p_{T_{cut}}$ are shown in Fig. \ref{fig:scal_cut}, and a
 significant contribution
from the scalars to the boosted Higgs signal is apparent for $p_{T_{cut}} \sim M_S$ for $M_S=500$~\GeV.
For the heavier scalar, $M_S\sim $~\TeV, the effects are only a few $\%$ even for very large $p_{T{cut}}$.

\begin{figure}[t]
\begin{centering}
\includegraphics[scale=0.385]{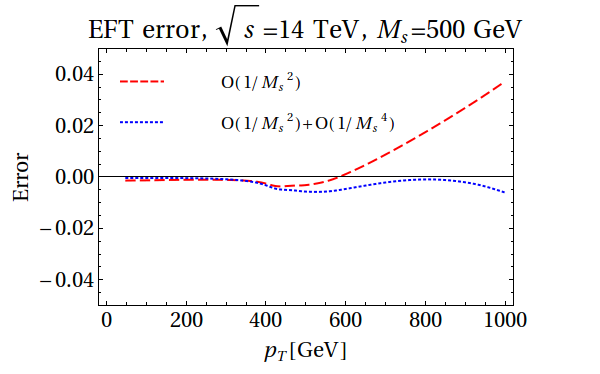}
\includegraphics[scale=.385]{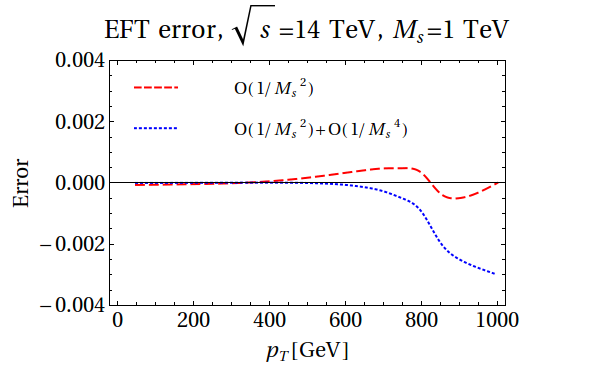}
\par\end{centering}
\caption{Accuracy of the effective field theory calculation of $d\sigma/ dp_T$
relative to the exact calculation when including $500 ~\GeV$ (LHS) and $1~\TeV$ (RHS) color triplet
scalars including all partonic initial states. 
The dashed lines contain only the dimension-$5$ contributions, while the dotted lines contain both
the dimension-5 and dimension-$7$ contributions. The SM top quark contribution is always included exactly.}
\label{fig:eft_tot}
\end{figure}

\begin{figure}[t]
\begin{centering}
\includegraphics[scale=0.385]{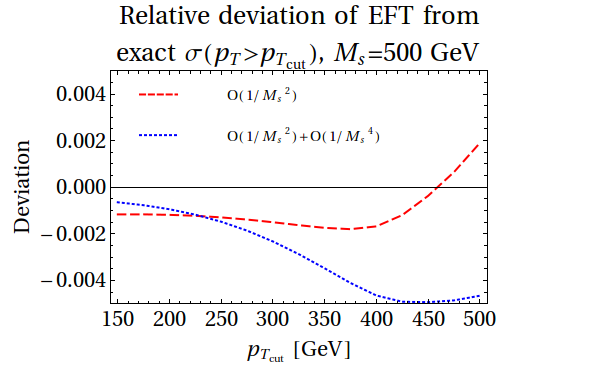}
\includegraphics[scale=.385]{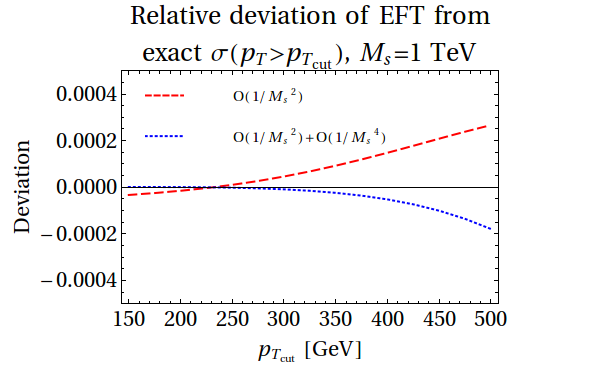}
\par\end{centering}
\caption{Accuracy of the effective field theory calculation of the total cross section
subject to a $p_{T_{cut}}$,
relative to the exact calculation when including $500 ~\GeV$ (LHS) and $1~\TeV$ (RHS) color triplet
scalars including all partonic initial states. 
The dashed  lines contain only the dimension-5 contributions, while the dotted lines contain both
the dimension-5 and dimension-7 contributions. The SM top quark contribution is included exactly.}
\label{fig:eft_tot_int}
\end{figure}

\begin{figure}[t]
\begin{centering}
\includegraphics[scale=0.38]{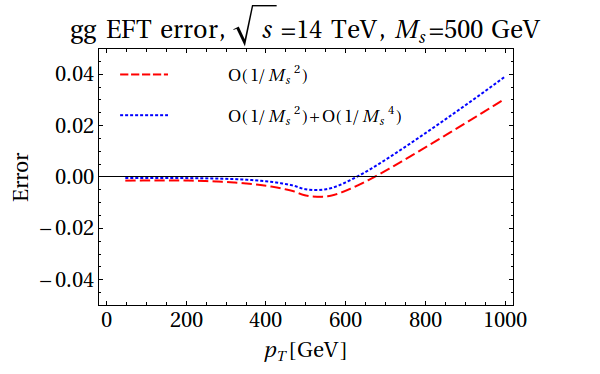}
\includegraphics[scale=.38]{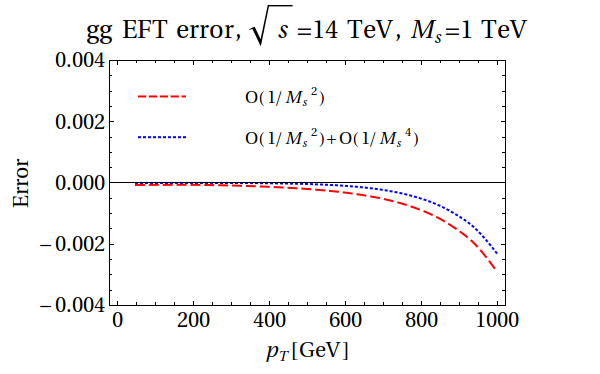}
\par\end{centering}
\caption{Accuracy of the effective field theory calculation of $d\sigma/ dp_T$
relative to the exact calculation when including $500 ~\GeV$ (LHS) and $1~\TeV$ (RHS) color triplet
scalars and including only the $gg$ initial state. The dashed  lines contain only the dimension-5 contributions, while the dotted lines contain both
the dimension-$5$ and dimension-$7$ contributions. The SM top quark contribution is included exactly.}
\label{fig:eftgg}
\end{figure}

\begin{figure}[t]
\begin{centering}
\includegraphics[scale=0.38]{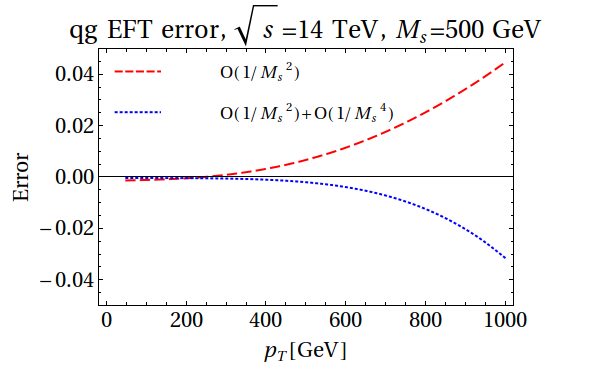}
\includegraphics[scale=.38]{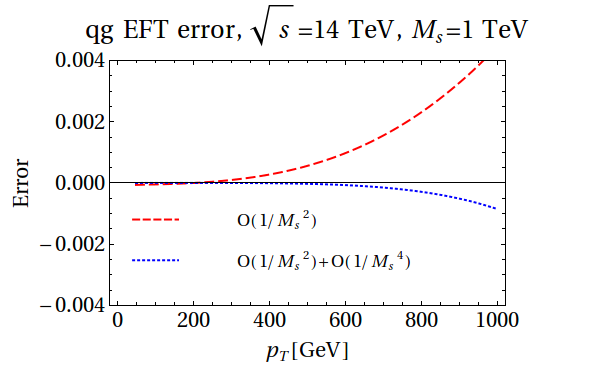}
\par\end{centering}
\caption{Accuracy of the effective field theory calculation of $d\sigma/ dp_T$
relative to the exact calculation when including $500 ~\GeV$ (LHS) and $1~\TeV$ (RHS) color triplet
scalars and including only the $qg$ initial state. The dashed  lines contain only the dimension-5 contributions, while the dotted lines contain both
the dimension-$5$ and dimension-$7$ contributions. The SM top quark contribution is included exactly.}
\label{fig:eftqg}
\end{figure}

Since the lowest order contribution from scalars is known exactly, 
we can explore the range of validity of the EFT.
Fig. \ref{fig:eft_tot} 
shows the deviation of the EFT calculation from the exact 1-loop result
when color triplet scalars are included.  For a $500~\GeV$ scalar, the EFT is accurate
to within a few $\%$ below $M_S$ and 
has large deviations above $500~\GeV$ when only the dimension-$5$ ($\sim 1/ M_S^2$) contributions
are included.  Including the dimension-7 contributions improves the accuracy of the EFT.  Interestingly,
for $M_S=1~\TeV$, the EFT becomes less accurate at large $p_T$ when the dimension-$7$ effects are included.  
The EFT expansion clearly breaks down at a scale $p_T\sim M_S$. 
Fig. \ref{fig:eft_tot_int} demonstrates the accuracy of the EFT in the $p_T$ integrated cross section and
we observe the same behavior.  (The cross section is integrated to $p_T=1~\TeV$, where the EFT is breaking
down.  Since the partonic results are integrated with a falling PDF spectrum, we expect the results to
be reasonably reliable.)
The contributions from the $gg$ and $qg$ initial states are shown separately 
in Figs. \ref{fig:eftgg} and \ref{fig:eftqg}.

  \subsection{Heavy Fermion Top Partners} 
\begin{figure}[t]
\begin{centering}
\includegraphics[scale=.36]{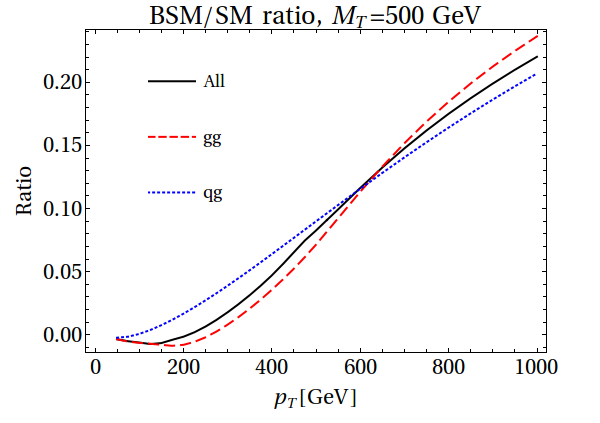}
\includegraphics[scale=.36]{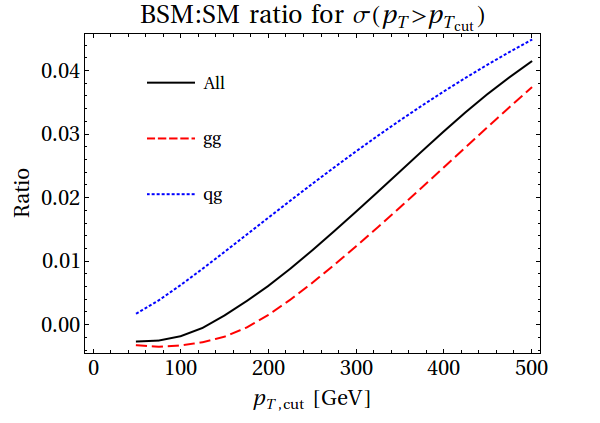}
\par\end{centering}
\caption{The BSM contribution, relative to the SM contribution, to the differential (LHS) and integrated Higgs $p_T$ distribution (RHS). The $gg$ and $qg$ partonic channels, and the sum of all partonic channels (which also include $q\bar q$), are shown separately. Both the top partner and top quark contributions are included exactly at LO.}
\label{fig:efttop}
\end{figure}

\begin{figure}[t]
\begin{centering}
\includegraphics[scale=.38]{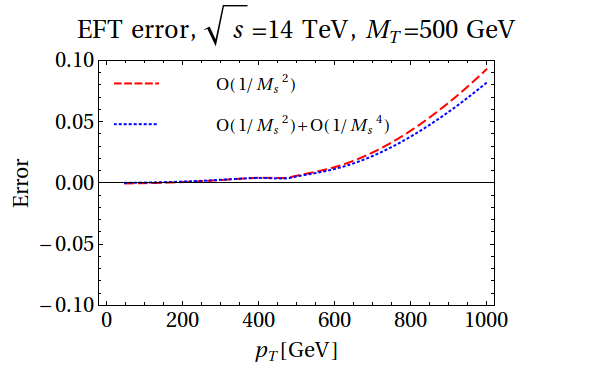}
\includegraphics[scale=.38]{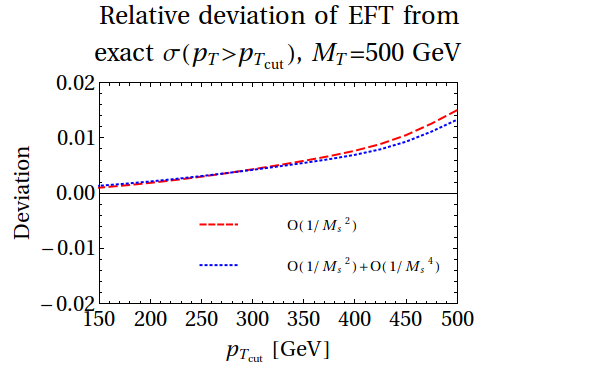}
\par\end{centering}
\caption{Accuracy of the effective field theory calculation of the differential (LHS) and integrated (RHS) $p_T$ distribution,
relative to the exact calculation, for a 500 GeV fermionic top partner with $\theta = \pi / 12$. }
\label{fig:efttop2}
\end{figure}

  In this section we consider the effect of a top partner model on the shape of the Higgs $p_T$ distribution. 
  We take the top partner mass  $M_T=500~\GeV$ and the mixing angle $\cos\theta_L= 0.966$.
  Fig. \ref{fig:efttop} shows the ratio of  the inclusive cross section  in  the top
  partner model to that in the SM, minus 1, evaluated with the exact dependence on the masses $m_t$ and $M_T$,
  along with the same quantity integrated with a $P_{T_{cut}}$.
  We note that the results of Ref. \cite{Banfi:2013yoa} demonstrate large effects at high 
  $p_T\sim1$~\TeV\  when $\sin\theta_L=0.4$. 
  Regretably, such large mixing angles are excluded by precision electroweak data.  (We agree with the 
  results of Ref. \cite{Banfi:2013yoa} for small $\sin\theta_L$.)
  Fig. \ref{fig:efttop2} shows the accuracy of the EFT predictions for differential and integrated $p_T$ distributions, relative to the results with exact $m_t$ and $M_T$ dependence.

\begin{figure}[t]
\begin{centering}
\includegraphics[scale=0.38]{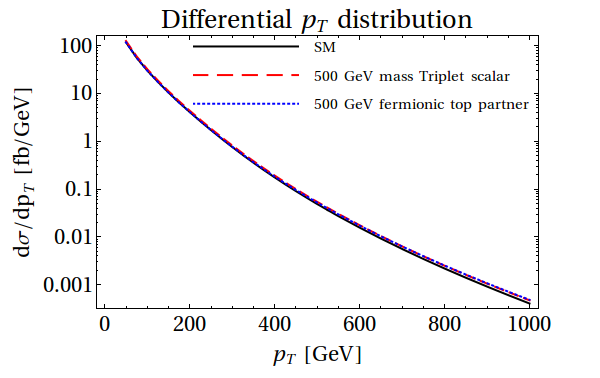}
\includegraphics[scale=.38]{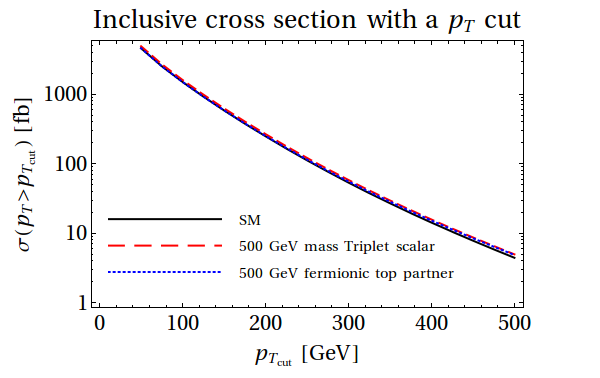}
\par\end{centering}
\caption{Cross sections including  the SM result and  a $500~\GeV$ ~color triplet scalar , the SM result and a  $500$~\GeV~ top
partner, compared with the SM predictions.}
  \label{fig:sum}
\end{figure}

We close this section by summarizing our results for top partners and scalars in Fig. \ref{fig:sum},
which dramatically demonstrates the difficulty  of extracting information about the underlying UV
physics.

  \subsection{EFT Fits}
\begin{figure}[t]
\begin{centering}
\includegraphics[scale=0.3]{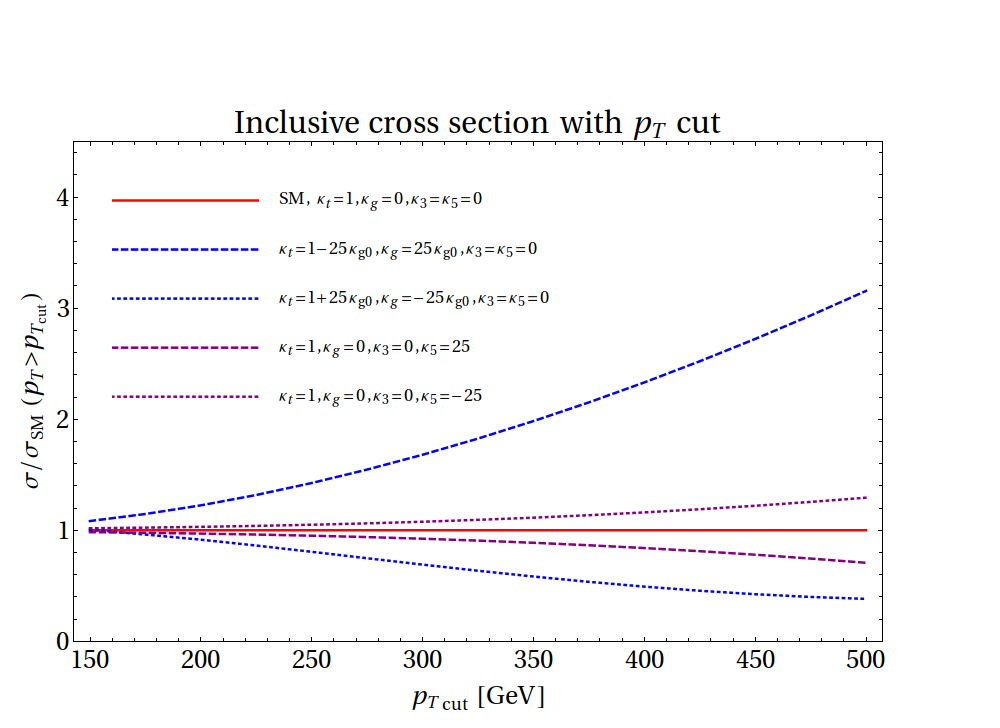}
\par\end{centering}
\caption{Inclusive cross section with a $p_T$ cut at $\sqrt{ s}=14$~\TeV, normalized to the SM rate. In our parameterization of BSM effects, the SM rate is rescaled by $\kappa_t$, while $C_1$, $C_3$, and $C_5$ are rescaled by $\kappa_g$, $\kappa_3$, and $\kappa_5$, respectively, with the model in Subsection \ref{subsec:scalar} corresponding to $|\kappa_g| / \kappa_{g0}=\kappa_3=\kappa_5 = 1$, $\kappa_{g0} \approx 0.0337$. We have fixed $\kappa_t+\kappa_g = 1$ to approximately conserve the total cross section. $\kappa_3$ is fixed to zero in this plot to highlight the effects of $\kappa_g$ and $\kappa_5$.}
\label{fig:eft_gen}
\end{figure}

\begin{figure}[t]
\begin{centering}
\includegraphics[scale=0.3]{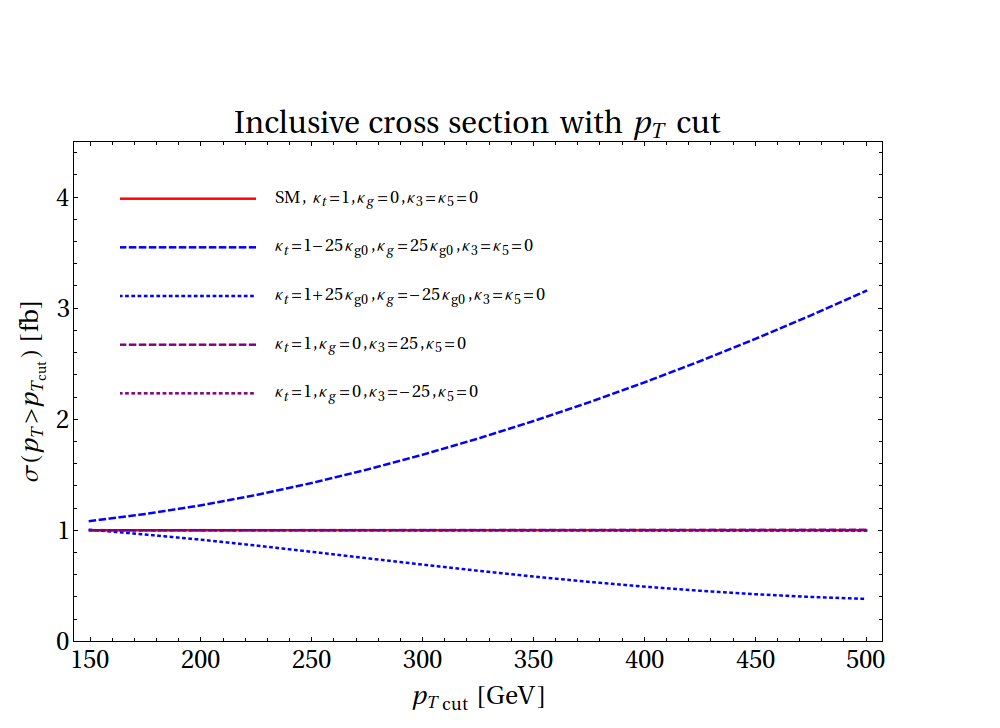}
\par\end{centering}
\caption{Inclusive cross section with a $p_T$ cut at $\sqrt{ s}=14$~\TeV, normalized to the SM rate. In our parameterization of BSM effects, the SM rate is rescaled by $\kappa_t$, while $C_1$, $C_3$, and $C_5$ are rescaled by $\kappa_g$, $\kappa_3$, and $\kappa_5$, respectively, with the model in Subsection \ref{subsec:scalar} corresponding to $|\kappa_g| / \kappa_{g0}=\kappa_3=\kappa_5 = 1$, $\kappa_{g0} \approx 0.0337$. We have fixed $\kappa_t+\kappa_g = 1$ to approximately conserve the total cross section. $\kappa_5$ is fixed to zero in this plot to highlight the effects of $\kappa_g$ and $\kappa_3$. The effect of $\kappa_3$ can be seen to be extremely small.}
\label{fig:O3}
\end{figure}

  In this subsection, we consider the effects of a general rescaling of the EFT coefficients.  As in Eq. \eqref{eq:kappaF} and Eq. \eqref{eq:defKg}, we consider the SM top quark contribution rescaled by $\kappa_t$, and the $C_1$ coefficients rescaled by $\kappa_g$ relative to an infinitely heavy Dirac fermion whose mass comes entirely from the Higgs, i.e. $C_1 = \kappa_g \cdot \alpha_s / (12\pi v)$. For the dimension-7 operators, we vary the matching coefficients $C_i =  \kappa_i C_i(M_S=500~{\rm GeV}, C_h =3m_Z)$ for $i=3,5$,
  where the reference values, scaled by $\kappa_i$, are $C_3(M_S, C_h)=-g_s \alpha_s C_h / (1440 M_S^4)$ and $C_5(M_S, C_h) = -\alpha_s C_h / (360 \pi M_S^4)$  corresponding to the EFT coefficients from Table I for a $500$~\GeV \ scalar. The total cross section for single Higgs production is roughly unchanged from the SM, if we fix $\kappa_t+\kappa_g$ to be 1, according to Eq. \eqref{eq:SMrel}. Fig. \ref{fig:eft_gen}
   demonstrates that excessively large
  values of $\kappa_5$ are required for a large effect from $O_5$. Fig. \ref{fig:O3} shows that the inclusion of $O_3$ has very little effect even for huge values of $\kappa_3$, as expected from the helicity arguments in \cite{Dawson:2014ora}.  On the other hand, the effect of rescaling $\kappa_t$ and $\kappa_g$ separately
  can have a relatively large effect.

\section{Conclusion}
\makeatletter{}The process Higgs + jet has been proposed as a useful channel for studying BSM physics 
and for disentangling the effects of a modification of the dimension-$4$ $t {\overline t}h$ Yukawa
coupling from a non-SM dimension-$5$ Higgs-gluon effective vertex.  We  further include dimension-$7$
effective Higgs-gluon operators and compute the EFT coefficient functions in two representative 
models with heavy colored scalars and fermions.  The coefficient functions are suppressed
by inverse powers of the heavy mass scales, $m$, and are therefore quite small.

At lowest order, the effects of colored scalars and fermions can be computed exactly and
the accuracy of the EFT determined.  Typically, better accuracy is obtained in the $gg$
channel than in the $qg$ channel, and the EFT is accurate to a few percent for $p_T < m$.
Our results illustrate the dilemma of the EFT approach:  large effects are only obtained at
high $p_T$ and the contribution from the dimension -$7$ operators is small for $p_T < m$.  On
the other hand, Fig. \ref{fig:eft_gen} demonstrates a modest sensitivity to $C_1$, independent of $\kappa_t$.
If any deviation is found in the Higgs transverse momentum distribution up to $1~\TeV$, the deviation is unlikely to provide
information about the UV physics beyond the single parameter $C_1$. Our results support the validity of an approach using only the dimension-$5$ Higgs-gluon operator.
 Inclusion of the NLO QCD corrections is unlikely to change this conclusion, since the NLO corrections to the $C_1^2$ contribution do not have a large $p_T$
 dependence in the region where the EFT is valid. 
 
\label{conc}

\begin{acknowledgments}
S.D. thanks A. Ismail and I. Low for discussions about the effects of virtual scalar particles. 
The work of SD and IL is supported  the U.S. Department of Energy under grants
No.~DE-SC0012704 and DE-AC02-76SF00515.  The work of MZ is supported by NSF grant PHY-1316617.
\end{acknowledgments}
\clearpage
\bibliography{hjet}

\end{document}